\documentclass{article}
\usepackage{spconf,amsmath,graphicx}
\usepackage{bm}
\usepackage{SIunits}
\usepackage{todonotes}
\usepackage{multirow}
\usepackage[acronym]{glossaries} 

\newcommand{\figref}[1]{Fig.~\ref{#1}}
\newcommand{\tableref}[1]{Table~\ref{#1}}
\let\originaleqref=\eqref
\renewcommand{\eqref}{Eq.~\originaleqref}

\ninept

\makeglossaries 
\newacronym{itd}{ITD}{interaural time difference}
\newacronym{ild}{ILD}{interaural level difference}
\newacronym{snr}{SNR}{signal-to-noise ratio}
\newacronym{erb}{ERB}{equivalent rectangular bandwidth}
\newacronym{dnn}{DNN}{deep neural network}
\newacronym{mfcc}{MFCCs}{mel-frequency cepstral coefficients}
\newacronym{drr}{DRR}{direct-to-reverberant ratio}
\newacronym{hats}{HATS}{head and torso simulator}
\newglossaryentry{brir}{name={BRIR},description={binaural room impulse response},
first={\glsentrydesc{brir} (\glsentrytext{brir})},
plural={BRIRs},
descriptionplural={binaural room impulse responses},
firstplural={\glsentrydescplural{brir} (\glsentryplural{brir})}}
\newglossaryentry{hrtf}{name={HRTF},description={head related transfer function},
first={\glsentrydesc{hrtf} (\glsentrytext{hrtf})},
plural={HRTFs},
descriptionplural={head related transfer functions},
firstplural={\glsentrydescplural{hrtf} (\glsentryplural{hrtf})}}
\newglossaryentry{hrir}{name={HRIR},description={head related impulse response},
first={\glsentrydesc{hrir} (\glsentrytext{hrir})},
plural={HRIRs},
descriptionplural={head related impulse responses},
firstplural={\glsentrydescplural{hrir} (\glsentryplural{hrir})}}
\newacronym{psd}{PSD}{power spectral density}
\newacronym{gmm}{GMM}{Gaussian mixture model}
\newacronym{kemar}{KEMAR}{Knowles Electronic Manikin for Acoustic Research}
\newacronym{rms}{RMS}{root mean square}
\newacronym{rmse}{RMSE}{root mean square error}
\newacronym{pdf}{PDF}{probability density function}
\newacronym{dft}{DFT}{discrete Fourier transform}
\newacronym{stft}{STFT}{short-time discrete Fourier transform}
\newacronym{em}{EM}{expectation-maximisation}
\newacronym{mct}{MCT}{multi-conditional training}
\newacronym{maa}{MAA}{minimum audible angle}
\newacronym{cnn}{CNN}{convolutional neural network}
\newacronym{tdoa}{TDOA}{time difference of arrival}
\newacronym{doa}{DOA}{direction of arrival}
\newacronym{srp}{SRP}{steered response power}
\newacronym{phat}{PHAT}{phase transform}
\newacronym{gcc}{GCC}{generalised cross-correlation}
\newacronym{ccf}{CCF}{cross-correlation function}
\newacronym{asr}{ASR}{automatic speech recognition}
\newacronym{mlp}{MLP}{multi-layer perceptron}

\title{END-TO-END BINAURAL SOUND LOCALISATION FROM THE RAW WAVEFORM}
\name{Paolo Vecchiotti\,$^1$, Ning Ma\,$^2$, Stefano Squartini\,$^1$ and Guy J. Brown\,$^2$\thanks{This work was carried out while P. Vecchiotti was visiting the University of Sheffield funded by an Erasmus+ Traineeship.}}
\address{
  $^1$\,Department of Information Engineering, Universit\`{a} Politecnica delle Marche, Ancona, Italy \\
  $^2$\,Department of Computer Science, University of Sheffield, Sheffield S1 4DP, UK \\ 
  {\small \tt p.vecchiotti@pm.univpm.it, \{n.ma, g.j.brown\}@sheffield.ac.uk, s.squartini@univpm.it}
}
\begin{document}
\ninept
\maketitle
\begin{abstract}

A novel end-to-end binaural sound localisation approach is proposed which estimates the azimuth of a sound source directly from the waveform. Instead of employing hand-crafted features commonly employed for binaural sound localisation, such as the interaural time and level difference, our end-to-end system approach uses a convolutional neural network (CNN) to extract specific features from the waveform that are suitable for localisation. Two systems are proposed which differ in the initial frequency analysis stage. The first system is auditory-inspired and makes use of a gammatone filtering layer, while the second system is fully data-driven and exploits a trainable convolutional layer to perform frequency analysis. In both systems, a set of dedicated convolutional kernels are then employed to search for specific localisation cues, which are coupled with a localisation stage using fully connected layers. Localisation experiments using binaural simulation in both anechoic and reverberant environments show that the proposed systems outperform a state-of-the-art deep neural network system. Furthermore, our investigation of the frequency analysis stage in the second system suggests that the CNN is able to exploit different frequency bands for localisation  according to the characteristics of the reverberant environment.
\end{abstract}

\section{Introduction}
\label{s:intro}

In the last decade, much effort has been spent towards the development of sound localisation systems. Classical approaches include estimation of the \glspl{tdoa} between microphone pairs using the \gls{gcc}~\cite{Knapp1976, brandstein1997practical}, beamformer based models such as SRP-PHAT~\cite{dibiase2000high}, and spectral estimation-based methods such as the multiple signal classification algorithm (MUSIC) \cite{schmidt1986multiple}.
More recently, localisation systems based on \glspl{dnn} have shown promising performance. In \cite{sun2018indoor}, probabilistic neural networks were used to estimate the \gls{doa} in an indoor environment using \gls{gcc}-based features. A similar scenario was studied in \cite{vesperini2018localizing} which used a \gls{cnn} to predict speaker coordinates. Binaural cues are employed in \cite{MaEtAl2017dnn}, where the \gls{ccf} was used as features in a \gls{dnn} to estimate the azimuth of a sound source with simulated head movement. \gls{cnn} architectures were also used in \cite{chakrabarty2017multi, adavanne2018sound} using frequency-domain features such as the phase or the magnitude of the signal.

All of the approaches so far are based on hand-crafted features explicitly extracted from the waveform. Such a feature extraction process may lead to a loss of information which can affect the performance. Human listeners, on the other hand, are able to use waveforms from just two ears to reliably determine the location of a sound source~\cite{Blauert97}. It is well known that this ability is largely based on both binaural cues, such as the \gls{itd} and the \gls{ild}, and monaural spectral cues created by direction-dependent filtering of the outer ears. However, it is less clear how these cues are seamlessly combined and processed by the auditory cortex for sound localisation~\cite{GrotheEtAl2010}.

Recently, much effort has been spent in the development of end-to-end systems for many audio applications. For example, a model for end-to-end \gls{asr} is proposed in \cite{sainath2017multichannel}, which combines localisation, beamforming, acoustic modelling and speech enhancement in a unified \gls{dnn}. In audio generation, several end-to-end methods were proposed to directly generate waveforms from text~\cite{van2016wavenet, mehri2016samplernn}. 

This paper proposes a novel end-to-end approach for sound localisation, referred to as \textit{WaveLoc}.
Instead of an explicit feature extraction stage, the proposed approach uses a \gls{cnn} with a cascade of convolutional layers to implicitly extract features directly from the raw waveform for sound localisation. One of the key stages in the network is the frequency analysis, and two different approaches are investigated. The first approach is auditory-inspired and uses a convolutional layer based on the gammatone filterbank~\cite{PattersonEtAl1992}. The gammatone filter is a widely-used model of auditory frequency analysis, with bandwidths set to reproduce human critical bandwidths~\cite{WangBrown2006}. In the second model, we adopt a standard convolutional layer which is intended to learn how to perform frequency analysis along with the training process of the entire network. After frequency analysis, further convolutional layers with 2-D kernels operates directly on the signals from both ears to extract features that are similar to the binaural cues used by the auditory system. The extracted features are finally concatenated and used as input to a \gls{dnn} with fully connected layers, in order to map them to the corresponding source azimuth.

Our evaluation shows that the proposed WaveLoc systems are able to accurately estimate the azimuth of a sound source in the anechoic condition. However, the performance of the data-driven WaveLoc approach is poor in reverberant conditions when trained only on anechoic signals. This leads to a detailed investigation of the benefits of \gls{mct}, following which we are able to demonstrate robust performance of the wave-based approaches across a range of challenging reverberant conditions. 

The rest of the paper is organised as follows. Section~\ref{s:system} presents the proposed end-to-end sound localisation framework, with a focus on two waveform-based approaches that differ in the frequency analysis stage. The experiment setups are described in Section~\ref{s:eval} and results are presented with discussions in Section~\ref{s:results}. Finally, Section~\ref{s:conc} concludes the paper and makes suggestions for future work.

\begin{figure*}[!h]
\centerline{\includegraphics[width=.9\textwidth]{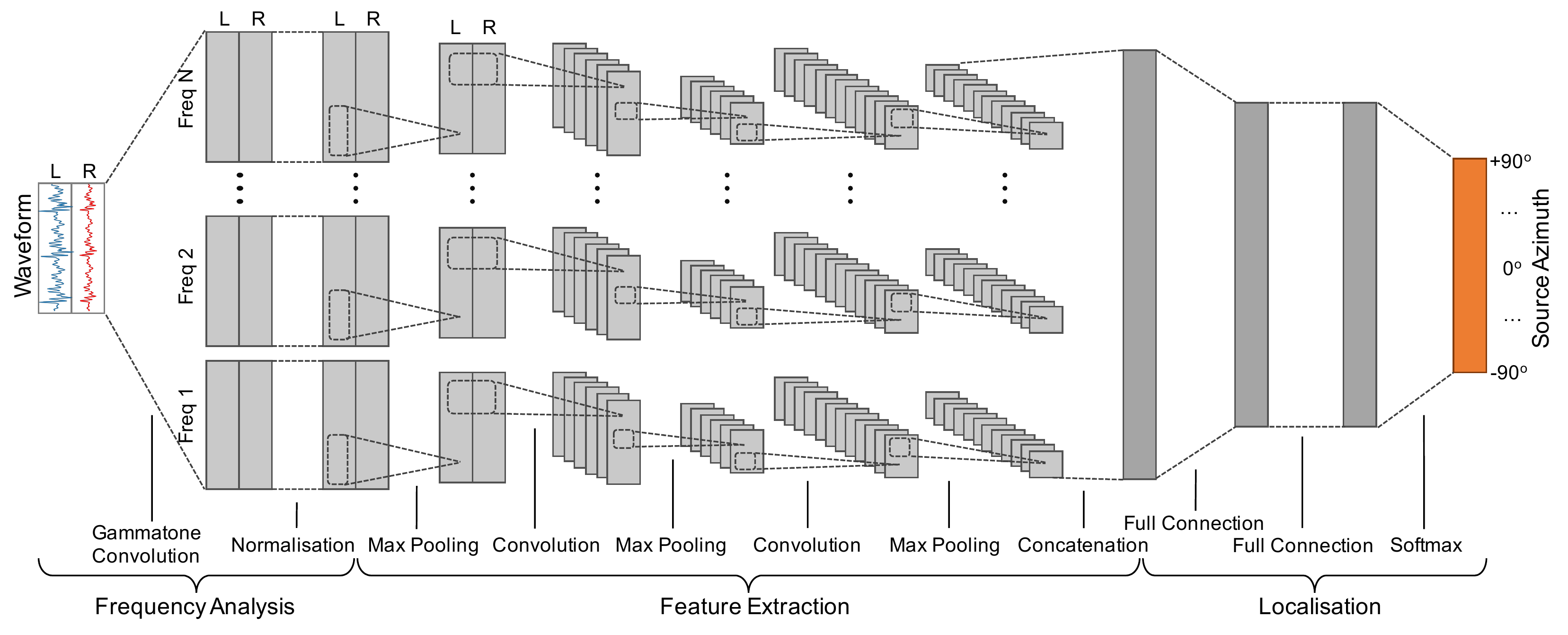}}
\caption{The proposed end-to-end WaveLoc-GTF system using convolutional neural networks for binaural sound localisation. }
\label{f:system}
\end{figure*}

\section{System Description}
\label{s:system}

\subsection{Overview}
\label{ssec:overview}

The proposed end-to-end sound localisation approach is illustrated in Fig.~\ref{f:system}. The convolutional neural network can be broadly divided into three stages: (i) a frequency analysis stage that takes the framed binaural ear signals as input, (ii) a feature extraction stage with a cascade of convolutional layers to extract suitable features for sound localisation, and (iii) a sound localisation stage based on several dense layers to perform sound localisation as a classification task.

The raw waveforms of the left and right ear signals, as indicated by `L' and `R' in Fig.~\ref{f:system}, are directly used as inputs to the proposed \glspl{cnn}. The ear signals are sampled at 16\,kHz and framed with 20\,ms window size with 10\,ms overlap. In each frame the left and right channels are stacked together to form an input matrix of size $2\times320$. 

It is well established that the auditory system performs a frequency analysis that divides the ear signal into frequency bands, and then does analysis on the fine time signal in each band~\cite{ehret1978stiffness, Yost2013}. Such processing has been shown to improve the robustness when exploited in a binaural sound localisation system, particularly in reverberant environments~\cite{MaEtAl2018loc}. To simulate this operation, the first stage of the \gls{cnn} performs a frequency analysis which filters the ear signals in the time domain with convolutional kernels.

Two frequency analysis strategies were investigated in this study. In the first system, named \textit{WaveLoc-GTF}, the frequency analysis is performed by a convolution layer which is broadly based on a gammatone filterbank~\cite{PattersonEtAl1992}. As shown in Fig.~\ref{f:system}, the frequency analysis layer consists of a number of frequency channels. The following layers in each frequency channel elaborate upon the frequency analysis output, in order to extract frequency-dependent features. The second system, named \textit{WaveLoc-CONV} imposes no constraint on frequency analysis. Instead, a convolutional layer with 1-D convolutional kernels is exploited to analyse frequency, with parameters learned from the data as part of the network training process.

In both systems, the frequency analysis is followed by a layer of 2-D convolutional kernels to extract features based on correlations of the left and the right channels. In WaveLoc-GTF these kernels are applied separately for each output of the gammatone filters, while in WaveLoc-CONV they are applied to the single frequency analysis layer. The correlation-based features are closely related to  \gls{itd} and  \gls{ild} cues, which are further elaborated by another convolutional layer with 1-D kernels in order to search for specific patterns that are related to the localisation task. Finally, the features produced by the convolutional layers are flattened and concatenated, before being passed to two dense layers. A softmax activation function is used in the output layer in order to perform sound localisation as a classification task.

\subsection{WaveLoc-GTF}
\label{ssec:waveloc-gtf}

Fig.~\ref{f:system} illustrates the first proposed \gls{cnn}: WaveLoc-GTF. As discussed, the frequency analysis is performed by a gammatone filter bank, which consists of 32 filters spanning between 70 and 7000\,Hz with peak gain set to 0\,dB. These filters are directly coded into \textit{non-trainable} \gls{cnn} kernels of size $1\times320$, with a linear activation function. The gammatone impulse response is given by:
\begin{equation}
w[t] = at^{n-1}\cos(2{\pi}ft+\phi)e^{-2\pi bt}
\end{equation}
where $t$ is time, $a$ is the amplitude, $f$ is the centre frequency, $\phi$ is the phase of the carrier, $n$ is the filter's order, and $b$ is the filter's bandwidth. In order to perform a time convolution, each filter is flipped in time so that the kernel operation is defined as:
\begin{equation}
y[t] = \sum_{ m = -M}^{M} x[m] w[t - m]
\end{equation}
where $x$ is the input signal, $w$ the weights of the filter, $t$ is the index of the actual value and $M$ is the filter length.


In each frequency band, the resulting feature maps share the same dimensions ($2\times320$) of the input matrix.  A normalisation layer is then applied which looks for the maximum absolute value across all the gammatone channels before dividing them by this value. Hence, the output feature values range between [-1,1], which are further processed with $1\times2$ max pooling.


A separate stack of two further convolutional layers processes each normalised channel, searching for specific patterns related to localisation. The first convolutional layer has 2-D kernels of size $2\times18$ and the second layer has a set of 1-D kernels of size $1\times6$. Both convolutional layers are followed by $1\times4$ max pooling and employ \textit{ReLU} activation. Finally, the processed channels are concatenated. and fed into two fully connected dense layers. 
Each dense layer consists of 1024 hidden units with \textit{ReLU} activation and a dropout rate of 0.5. 

The output layer consists of 37 nodes corresponding to the 37 azimuth classes, with \textit{softmax} activation.

\subsection{WaveLoc-CONV}

The neural architecture of the second system, WaveLoc-CONV, employs a single convolutional layer dedicated to frequency analysis. Its key difference from WaveLoc-GTF is that the frequency analysis of this model is learnt during the training process together with other parameters of the network. A convolutional layer with 64 1-D kernels of shape $1\times256$ is employed as time domain filters for frequency analysis. It is reasonable to expect that the shape of a convolutional kernel directly trained on a raw waveform will be similar to all the sinusoidal components that form the waveform itself. In other words, the convolutional kernels are characterised by a set of sinusoidal functions, which lead to a particular frequency response of the kernel itself~\cite{sainath2017multichannel}.


The convolutional layer is followed by $1\times2$ max pooling with a linear activation function applied. As in WaveLoc-GTF, two more convolutional layers are employed to search for features suitable for localisation. However, instead of acting separately for each channel as in WaveLoc-GTF, they now jointly process all the output of the frequency analysis stage. The first of the two layers uses 64 2-D kernels of size $2\times18$ to look for correlations between the left and right channels. The second uses 64 1-D kernels of size $1\times6$. Both layers use the ReLu activation function and are followed by $1\times4$ max pooling. Finally, the outputs are flattened and fed into a two fully-connected hidden layers with 1024 units each. The output layer uses softmax activation with 37 neurons.

The hyperparameters or both end-to-end architectures are chosen based on an optimisation process using a development dataset.

\section{Evaluation}
\label{s:eval}


\subsection{Binaural simulation}

Binaural signals were simulated by convolving speech recordings with the Surrey \gls{brir} database \cite{HummersoneMasonBrookes10}. The Surrey \glspl{brir} were captured using a Cortex \gls{hats} in both anechoic and reverberant rooms. A total of 37 azimuth angles were used, ranging from [-90\degree, 90\degree] in steps of 5\degree, where 0\degree \space is located exactly in front of the head.
Four reverberant rooms were employed, denoted A--D. The reverberation time (T$_{60}$) and \gls{drr} of each room are listed in \tableref{t:rooms}.

\begin{table}[thb]
\caption{Room characteristics of the Surrey \gls{brir} database~\cite{HummersoneMasonBrookes10}.}
\label{t:rooms}
\vspace{1mm}
\centering
\begin{tabular}{@{} l  c  c  c  c  @{}}
\hline\hline
 & Room A & Room B  & Room C & Room D \\
\hline
$\mathrm{T}_{60}$ (s) & 0.32 & 0.47 & 0.68 & 0.89  \\
$\mathrm{DRR}$ ($\deci\bel$) & 6.09 & 5.31 & 8.82 & 6.12 \\
\hline\hline
\end{tabular}
\end{table}

Speech signals belonging to the DARPA TIMIT database \cite{Gar1993} were convolved with each \glspl{brir}. The initial and final frames of each speech utterance were truncated if silence was present. The training dataset was obtained by randomly selecting 24 sentences per azimuth from the TIMIT training subset, while another 6 sentences composed the validation dataset. 15 more sentences per azimuth were selected from the TIMIT test subset to create the test dataset.

\subsection{Experimental setup}

For training the \textit{Adam} optimiser with a learning rate of $1\mathrm{e}{-3}$ and a batch size of 128 samples was employed. The training process lasted for 50 epochs, but early stopping was applied if no improvement was observed on the validation set for more than 5 epochs. A decreasing learning rate was employed to improve training, being multiplied by 0.2 if no lower error was achieved after 2 epochs.

The networks were trained in two acoustic room conditions: (i) using anechoic signals only for training; (ii) multiconditional training, in which the networks were trained using data from all the reverberant rooms apart from the one used for test.

The evaluation results are reported based on chunks. Each chunk is 250\,ms long (25 frames). The prediction made for each frame in a chunk is averaged to report a single azimuth location for the chunk. Chunk-based evaluation was adopted in order to avoid the issue that a speech signal typically includes short pauses where there is no directional sound source. The accuracy of the models was finally measured in terms of \gls{rmse} given in degrees.

\subsection{Baseline system}

The baseline system is a state-of-the-art \gls{dnn}-based localisation system using GCC-PHAT features as inputs \cite{vesperini2018localizing,xiao2015learning}. GCC-PHAT features are computed as the inverse transform of the frequency domain cross-correlation of two audio signals captured by a microphone pair.
The binaural signals sampled at 16\,kHz are framed at 20\,ms, with 10\,ms overlap. Since a distance of 18\,cm occurs between the two microphones, the first 37 values are selected from the inverse transform. Unit variance and zero mean normalization is then applied.
The baseline network consists of an input layer, two hidden layers of 1024 units each and an output layer of 37 classes. Dropout equal to 0.5 is applied after the two hidden layers. Softmax is selected as the activation function for the output layer, while a sigmoid activation function is used for the hidden units. All the hyperparameters were optimised using the development dataset.

\section{Results and Discussion}
\label{s:results}

\subsection{Anechoic training}
\label{ssec:anec_train}

\tableref{t:anechoic_train} shows results using systems trained in the anechoic condition. The best overall performance is achieved by the baseline GCC system. The proposed WaveLoc-GTF performed slightly worse compared to the baseline, while the localisation errors for WaveLoc-CONV were considerably larger across all reverberant conditions.

\begin{table}[!h]
\center
\caption{Localisation \gls{rmse} results in degrees for the models trained in anechoic environment.} 
\label{t:anechoic_train}
\vspace{1mm}
\begin{tabular}{@{} l c c c c c @{}}
\hline \hline
                                 Room                           & Anechoic &  A &  B &  C &  D \\ \hline
Baseline              & 0.1\degree        & \textbf{2.6\degree}    & \textbf{9.3\degree}   & 2.6\degree    & \textbf{10.1\degree}    \\ 
WaveLoc-GTF & \textbf{0\degree}        & 9.1\degree    & 10.7\degree    & \textbf{1.6\degree}    & 10.5\degree    \\ 
WaveLoc-CONV & \textbf{0\degree}       & 37.7\degree      & 41.8\degree    & 37.3\degree      & 44.4\degree      \\
\hline\hline
\end{tabular}
\end{table}

\begin{figure}[!h]
\center
\includegraphics[width=0.9\columnwidth]{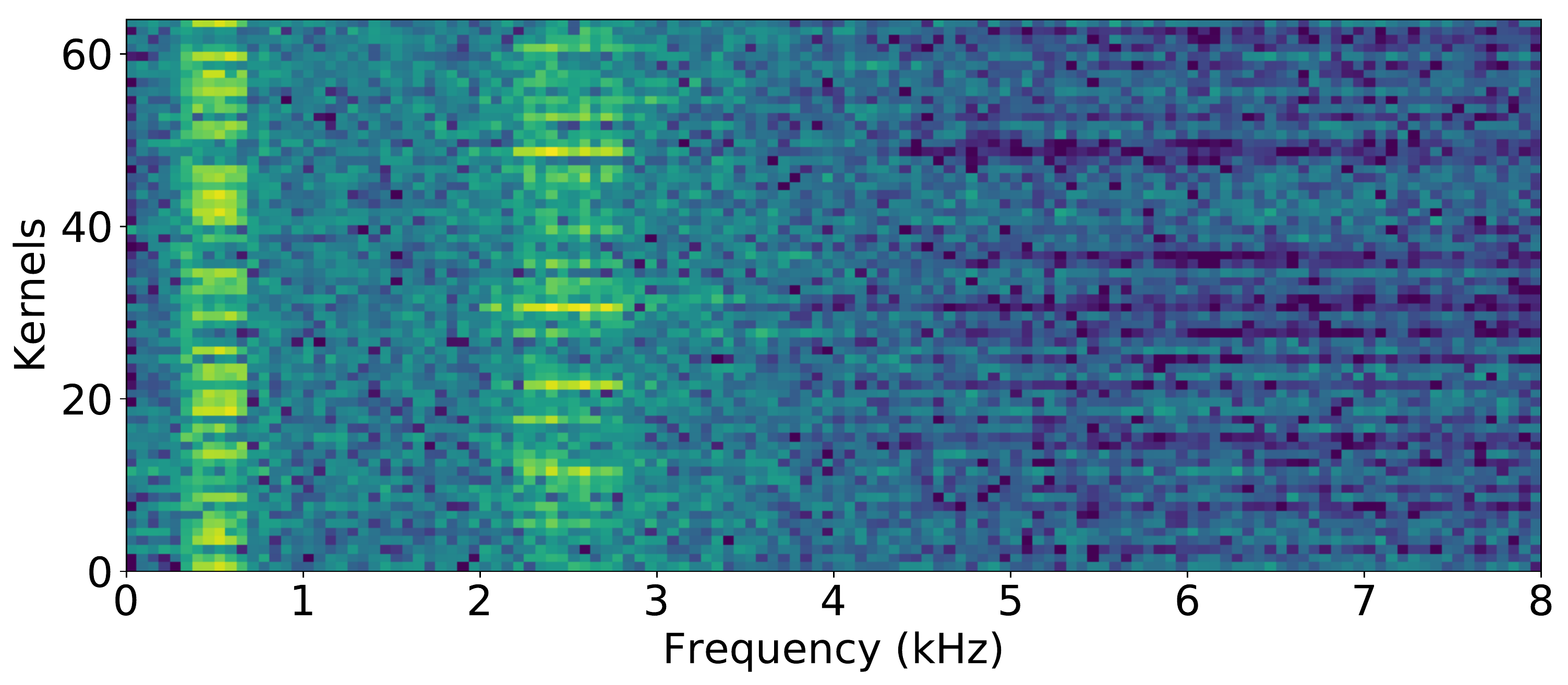}
\caption{Log-power spectra of the kernels in the first convolutional layer of WaveLoc-CONV when trained in the anechoic environment.}
\label{f:spec_train_anec}
\end{figure}

It appears that the WaveLoc-CONV system has a tendency for overfitting compared to the other two systems. \figref{f:spec_train_anec} shows the log-power spectra of all the 64 kernels in the first convolutional layer in WaveLoc-CONV. It is clear that the kernels, when trained in the anechoic condition, act largely as a set of band pass filters, mostly enhancing the frequency bands between 300--600\,Hz and between 2300--2800\,Hz. It is widely known that binaural features such as \glspl{itd} are more reliable in the low frequency region below 1600\,Hz while others such as \glspl{ild} become more robust in the high frequency region above 1600\,Hz~\cite{Blauert97}. It is possible that the network extracts related binaural features which are most effective in these two bands for sound localisation in the anechoic condition. Such behaviour, however, failed to generalise to unseen reverberant conditions as these frequency bands could become unreliable due to reverberation.

The WaveLoc-GTF model, on the other hand, performs frequency analysis with the gammatone filterbank layer which forces the system to exploit all frequency bands and thus extract the most effective localisation features in each band.

\subsection{Multiconditional training}
\label{ssec:mct_train}

It has been shown in the past that \gls{mct} can mitigate overfitting and increase the robustness of sound localisation in reverberant conditions~\cite{MaEtAl2017dnn, MayvandeParKohlrausch11a}. This can be done by adding either diffuse noise or reverberation to the training signals. In this study, a reverberant training approach was adopted as our preliminary experiments showed it to be more effective. Specifically, the anechoic training dataset was supplemented with reverberant versions by convolving it with various \glspl{brir}. 
For each one of the four reverberant room under evaluation, all the remaining three were included for \gls{mct}.

\begin{table}[thb]
\center
\caption{Localisation \gls{rmse}  results in degrees using \gls{mct}.} 
\label{t:train_mct_diffuse_room}
\vspace{1mm}
\begin{tabular}{@{} l c c c c @{}}
\hline \hline
                                   Room                         &  A &  B &  C &  D \\ \hline
Baseline           & 2.7\degree        & 3.3\degree    & 3.1\degree   & 5.2\degree        \\ 
WaveLoc-GTF & \textbf{1.5\degree}        & 3.0\degree    & 1.7\degree    & 3.5\degree        \\
WaveLoc-CONV & 1.7\degree       & \textbf{2.3\degree}      & \textbf{1.4\degree}    & \textbf{2.4\degree}         \\
\hline \hline
\end{tabular}
\end{table}

\tableref{t:train_mct_diffuse_room} lists the results of all the models. The anechoic condition was excluded in this study, as all the models performed well even without \gls{mct}. All the models benefitted from \gls{mct}, especially the proposed WaveLoc models. The best overall performance in reverberant conditions is achieved by the WaveLoc-CONV model, which has an average localisation \gls{rmse} less than 3\degree compared to over 30\degree without \gls{mct}.

\begin{figure}[t]
\center
\includegraphics[width=0.9\columnwidth]{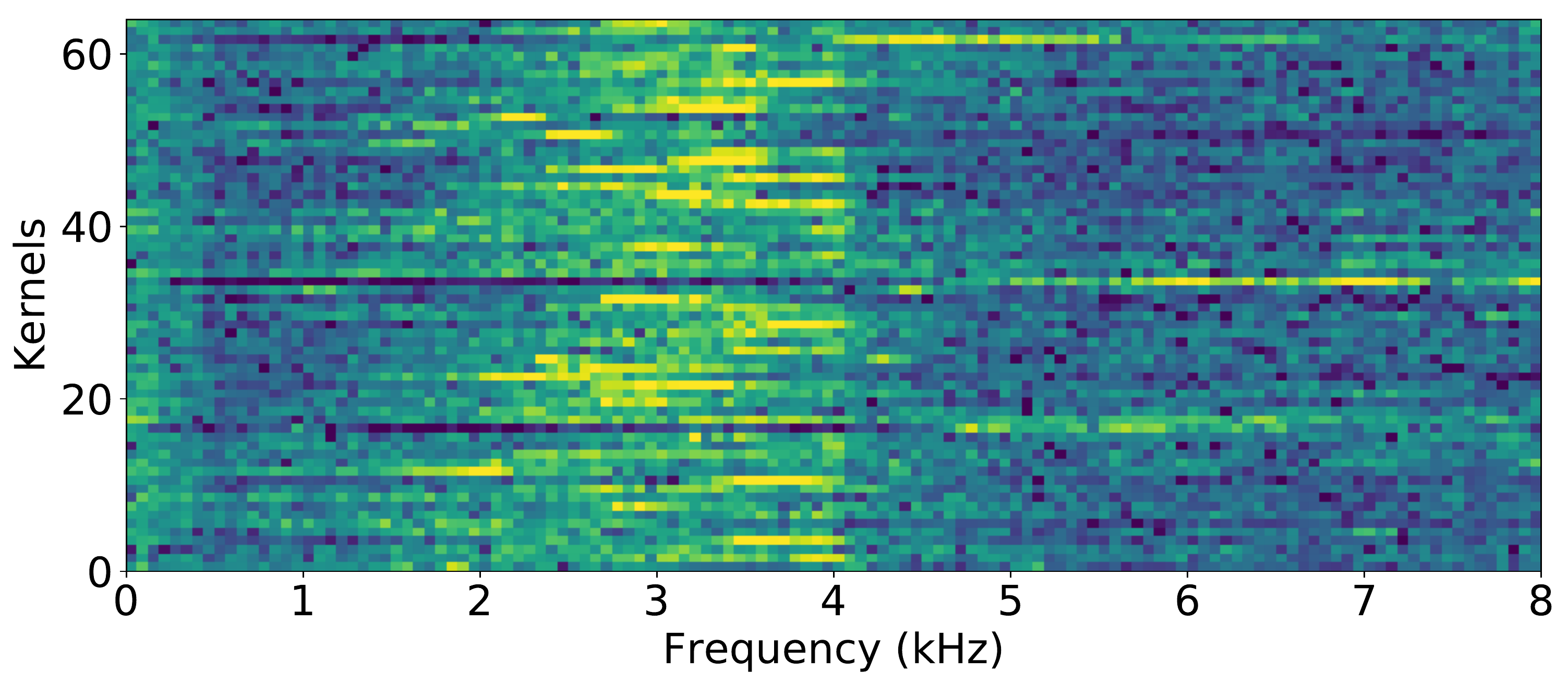}
\caption{Log-power spectra of the kernels in the first convolutional layer of WaveLoc-CONV when trained using \gls{mct}.}
\label{f:spec_train_mct_rooms_NO_D}
\end{figure}

To investigate the effect of \gls{mct} on the convolutional kernels, we again plot the log-power spectra of all the 64 kernels in the first convolutional layers of the WaveLoc-CONV model. Only the plot for Room D is shown in \figref{f:spec_train_mct_rooms_NO_D} but plots for all the other rooms are similar.
It can be seen that the first convolutional layer is now composed of a set of distributed bandpass filters emphasising mainly the 1500-4000\,Hz range, with some kernels stretching up to 6--7\,kHz.
The low frequencies below 1500\,Hz are less exploited by the WaveLoc-CONV model. It is interesting to notice that the data-driven model learns to use more high frequency cues in a reverberant environment, which suggests \gls{ild} become more useful than \gls{itd}. It is reasonable to expect that the \gls{itd} is more affected by reverberation, while the \gls{ild}, created by the head shadowing effect mainly for frequencies higher than 1600\,Hz, is more robust to reverberation. Indeed, psychophysical cue-trading studies find that human listeners give \gls{ild} more weight than \gls{itd} when localising sounds in reverberant conditions \cite{nguyen2014}.

\section{Conclusions}
\label{s:conc}

This paper described a new approach for localising a sound source directly from the waveform, by proposing two novel end-to-end \gls{cnn} systems.  Machine localisation systems typically employ hand-crafted features, such as the \gls{itd} and \gls{ild}. Such explicit feature extraction may limit the model performance since it implies a lossy transformation of the input signals. Instead, the proposed end-to-end approach employs a cascade of convolutional layers to extract  features directly from the waveform, that are suitable for localisation in reverberant environments. When \gls{mct} was used across reverberant conditions, both end-to-end systems outperformed a state-of-the-art \gls{dnn} system using conventional features.

Two \gls{cnn}-based systems were introduced. The first system, WaveLoc-GTF, is inspired by the auditory system and employs a convolutional layer that is largely based on a gammatone filterbank. The second system, WaveLoc-CONV, employs a data-driven approach, where a convolutional layer with trainable 1-D kernels is dedicated for frequency analysis. Although the gammatone filterbank is in some sense more `principled', since it approximates the filtering characteristics of the human auditory system, it does not work as well as a system that is trained (i.e., finds its own filters) across a number of reverberation conditions. One reason for this is that the system may elect to emphasise frequency regions during training that provide more robust cues to localisation. 

Indeed, we found that when \gls{mct} was used, the WaveLoc-CONV model was better able to exploit features in the high frequency regions above 2\,kHz, which tend to be less corrupted by reverberation. This mirrors findings from human perception suggesting that \gls{ild} (which is primarily available at high frequencies) is more robust than \gls{itd} when reverberation is present.

Future work will focus on improving the ability of end-to-end systems to generalise to unseen room conditions and multiple sources. Another possible direction is to combine sound identification with sound localisation within an end-to-end system~\cite{CakirVirtanen2018}. Finally, we plan to conduct `psychophysical' studies on trained networks in order to fully understand their underlying mechanisms, e.g. by using the cue trading protocol described in \cite{nguyen2014}. 

\vfill\pagebreak

\bibliographystyle{IEEEbib}
\bibliography{journal_abrv_short,deep_space}

\end{document}